\documentclass[12pt,letterpaper]{article}
\usepackage[utf8]{inputenc}
\usepackage{comment}
\usepackage{cite}
\usepackage[allow-number-unit-breaks]{siunitx}
\usepackage{amsmath,amssymb,amsfonts}
\usepackage{mathtools}
\usepackage{relsize}
\usepackage[framemethod=tikz]{mdframed}
\usepackage[toc,page]{appendix}
\usepackage{tabularx}
\usepackage{algorithmic}
\usepackage{graphicx}
\usepackage{textcomp}
\usepackage{xcolor}
\usepackage{xfrac}
\usepackage{lipsum}
\usepackage{adjustbox}
\usepackage{float}
\usepackage{placeins}
\usepackage{wrapfig}
\usepackage{ar}
\usepackage{caption}
\usepackage{subcaption}
\usepackage{url}
\usepackage{authblk}
\usepackage{tikz}       
\usepackage{pgfplots}
\pgfplotsset{compat=1.18} 
\usetikzlibrary{positioning, bending, shapes, fit, arrows, arrows.meta}

\begin{document}

\title{{\raggedright Self-Synchronized Trichel Pulse Trains in Multi-Point Corona Discharge Systems}}

\author[1]{Afshin Shaygani\thanks{Corresponding author: ashaygan@uwo.ca}}
\author[2]{Kazimierz Adamiak}

\affil[1,2]{Department of Electrical and Computer Engineering, Western University, London, Ontario, Canada N6A 5B9}
\affil[1]{ \texttt{ e-mail: ashaygan@uwo.ca}}

\renewcommand\Authands{ and }

\maketitle

\begin{abstract}

Evidence of self-synchronization has been observed in multi-electrode corona discharge systems, where the application of high negative DC voltages induces a self-sustained mode of current pulse trains. These pulses, historically referred to as Trichel pulses, characterize the operation of a two-electrode system where the discharge electrode is subjected to a high negative DC voltage. The numerical algorithm indicates that in a multi-electrode discharge system, comprising multiple two-electrode discharges, each two-electrode system independently produces pulse trains. These systems, each comprising a pair of electrodes, operate in a pulsed mode, with synchronization emerging from weak yet significant interactions among them. These interactions arise from the mutual influence of electric fields and space charges generated by each discharge pair. This influence extends beyond individual systems, leading to a synchronization between both pairs, both in a pulsed mode. A three-species model of discharge was employed to simulate this process and it was based on the finite element method formulation. Two different numerical models were investigated, a \(2D\) model, consisting of two discharge electrodes and a third grounded electrode, and two \(1D\)-axisymmetric models, consisting dual and triple pairs of discharge systems. Experiments show a multi-stable nature of the coupled pulsed discharge systems, indicating that under appropriate conditions the pulse trains exhibit two distinct modes of synchronization: in-phase synchronization and anti-phase synchronization. The occurrence of each mode depends on factors such as interaction strength, applied voltage level, and various system parameters. Furthermore, variations in these factors can lead to additional outcomes, including out of phase synchronization, as well as scenarios involving near-harmonic oscillations and quenching.

\end{abstract}

{\setlength{\parindent}{0cm}
}

\section{Introduction}

Synchronization is a phenomenon observed in self-sustained oscillators, where systems have a natural tendency to oscillate at a certain frequency without external influence. For synchronization to occur, there must be a weak coupling or interaction between the systems, allowing them to influence each other without forcing them into different behavior. The systems must also exhibit similarity in frequency or behavior, falling within a specific range of parameters that allow them to "lock" into a synchronized state.

The process of synchronization often requires gradual adjustment over time, where systems may drift in and out of synchronization due to factors like noise or slight changes in parameters. Additionally, synchronization can exhibit various modes, including in-phase and anti-phase synchronization, with the potential for more complex patterns. The interaction between these elements forms the basis for synchronization in various systems, from pendulum clocks to heartbeats and neuronal firing \cite{o2017oscillators,mirollo1990synchronization}.

One of the aspects of synchronization process is the dependence of the final synchronized state on the initial conditions or the initial phase difference between the systems. In basic synchronization models such as the Kuramoto model \cite{kuramoto1975self}, the final synchronized state (in-phase, out-of-phase, etc.) is independent of the initial phase difference. The initial state will affect how quickly they synchronize but not the final state. Conversely, in more complex systems or under certain conditions, the relationship between initial and final states may be more intricate, with factors such as network topology, coupling type, nonlinearity, and noise playing roles. For example, in multi-stable systems, there may be multiple synchronized states, and the initial phase difference can determine which state the system ultimately adopts. Thus, in certain contexts, the initial conditions are pivotal in shaping the final synchronized state, leading to complex synchronous behaviors 
\cite{books/cu/PikovskyRK01}.

The study of synchronization in coupled oscillators has been greatly influenced by the works of Kuramoto \cite{kuramoto1975self} and Winfree \cite{winfree1967biological,winfree1980geometry}. Research into this phenomenon spans a broad spectrum of systems. It encompasses simple models, such as those found in metronomes and clocks \cite{goldsztein2021synchronization,goldsztein2022coupled}, and extends to more complex and nonlinear frameworks including transistor-based chaotic oscillators \cite{minati2022incomplete,minati2022synchronization} and memristively-coupled oscillators \cite{ignatov2016synchronization}. The field also explores large-scale networks of oscillators \cite{leyva2013explosive,o2017oscillators,dai2020explosive}, and delves into more complicated nonlinear systems involving phenomena like chaos and turbulence \cite{herrmann2020modeling}. However, little exploration has been done to study this phenomenon in self-sustained electric discharges, such as corona discharge.

Corona discharge is a low-energy electric discharge that exhibits distinct modes based on polarity and voltage levels, all while being locally self-sustaining. Within certain voltage ranges and operating conditions, both negative and positive corona discharges can produce self-pulsations, in the negative discharge known as Trichel pulses \cite{trichel1938mechanism}. In Trichel pulse formation, applied voltages typically fall within the range of 
\({-5}\) to \(-\SI{50}{kV}\), and the pulses may operate in a quasi-steady-state regime with a frequency range from a few kilohertz to a few hundred kilohertz, and an average current ranging from \({5}-\SI{100}{\mu A}\). Such pulses are greatly influenced by factors like voltage level, temperature, pressure, humidity, and chemical compositions of the gas \cite{he2014spatio,xu2015influence,li2014influence,sattari2011numerical,sattari2011trichel,zhang2022effect,lama1974systematic}. This phenomenon is not exclusive to Trichel pulses, but has also been observed in various electrode configurations and geometries, including hollow cathode discharges (HCD) and parallel-plate configurations, elucidated in a concise comparative analysis between HCD, parallel-plate discharge pulses, and regular Trichel pulses generated in curved geometries \cite{xia2018comparison}.

The existence of seed electrons is vital in the development of self-sustained discharges, which can lead to either pulsed or pulse-less discharges. These seed electrons are generated either by photons reacting with the neutral molecules of air, inducing the photoionization effect (otherwise known as impact ionization), or as a result of the positive ions or high-energy photons striking the discharge electrode. The complex nature of discharges includes significant contributions from chemical reactions between charge species and the phenomena of charge transport influenced by electric fields, space charges, and neutral molecules. When dealing with two or more pulsed discharge systems interacting through these factors, it's logical to consider that their nonlinear dynamics could synchronize under the right conditions, such as weak coupling, allowing them to oscillate within the same frequency range without external influence. This intricate interplay between physical phenomena underscores the complexity and potential applications of multi-discharge systems in various scientific and industrial contexts, such as in designing ionic thrusters for advanced space propulsion systems, \cite{lozano2022densification,krejci2018space}, and electrospray techniques \cite{shirai2014atmospheric}.

Since Trichel's initial experiment \cite{trichel1938mechanism}, Trichel pulses have been the subject of extensive experimental study. Lama and Gallo's systematic research \cite{lama1974systematic} further illuminated the relationship between pulse frequency and discharge current, demonstrating how these aspects depend on both the applied voltage and the geometry of the two-electrode system. Earlier in their experimental study in 1973, the synchronization of Trichel pulses was initially observed, with the method detecting synchronization through current measurements rather than capturing the pulse waveforms \cite{lama1973interaction}. This subject remained largely unexplored until the presented numerical investigation, which not only confirmed the synchronization, but also captured the pulse waveforms. Moreover, this study revealed that synchronization can occur in more than two discharge systems and traced its origin to weak interactions stemming from the combined effects of space charges and electric fields. Building on this, the present work employs a three-species discharge model to delve into Trichel pulses dynamics in multi-electrode discharge systems, exploring the phenomenon of self-synchronization due to weak interactions of multiple discharges. The three-species discharge model involves the continuity of charge species, including chemical reactions, and the Poisson equation for the electric field. Considering a 2D model and two 1D-axisymmetric models for both two and three interacting discharge systems, the finite element method was employed for spatial discretization. The Trichel pulse trains were recorded and analyzed.

\section{Problem statement}

\subsection{Model specification}

\begin{figure}[htbp]
\centerline{\includegraphics[width=0.90\textwidth]{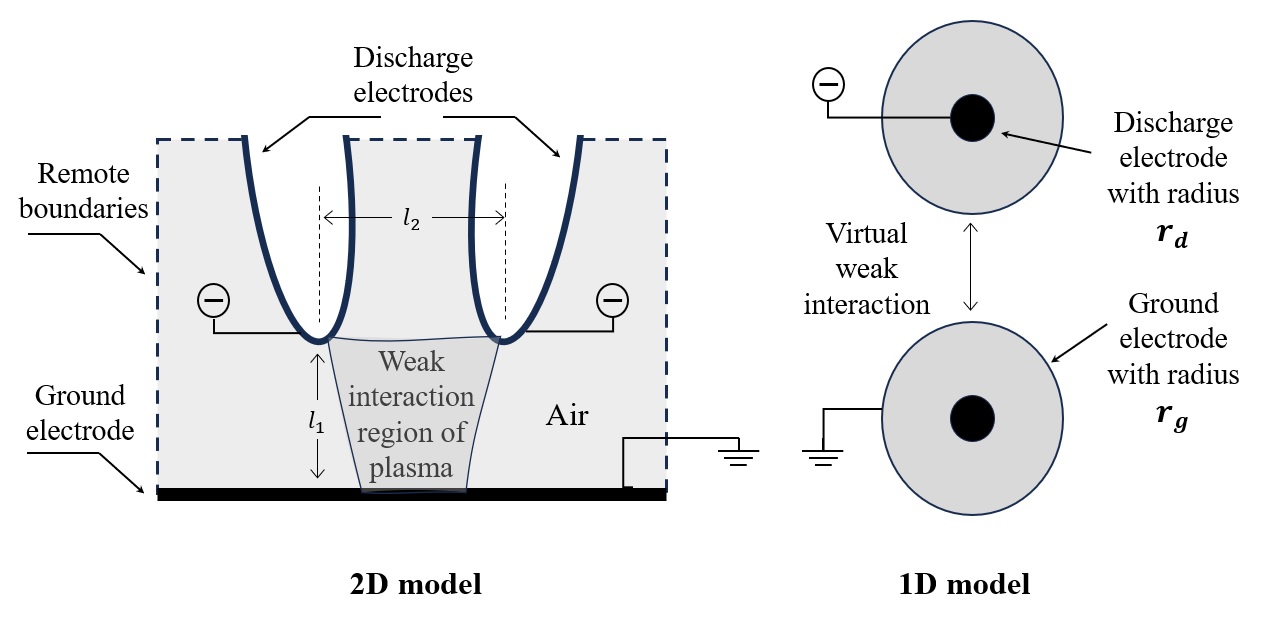}}
\caption{Illustration of a multi-electrode discharge system with weak plasma interactions, for \(2D\) and \(1D\)-axisymmetric models.}
\label{schematicE}
\end{figure}

The two models considered in this study are shown in  Fig. \ref{schematicE}. The discharge is in the Trichel pulse regime of the negative corona discharge. The \(2D\) model is a blade-plane type discharge with three electrodes: two blade electrodes having a hyperbolic profile, separation by \(l_2\), supplied with a high negative voltage \(V_{ap}\), and a ground electrode. The distance between the blades and the plate is \(l_1\). 

Two \(1D\)-axisymmetric models are considered, with a three and a two discharge systems. The schematic shows only the two electrode discharge system which comprises of two separate discharge systems of wire-cylinder type discharge, each embodying two electrodes: a wire electrode supplied with a high negative voltage \(V_{ap}\), and a cylinder which is grounded, with radii \(r_d\) and \(r_g\). 

Standard gas conditions are assumed for the simulations: the atmospheric pressure and temperature \(\SI{300}{K}\). The relative permittivity of the dry air filling the gap between the electrodes is \(\epsilon_r\).

\subsection{Governing equations of the \(2D\) model}

The mathematical representation of electric discharge considers  three fundamental charge components, including electrons, positive and negative ions \cite{shaygani2023numerical}. Three charge transport equations  govern the drift, diffusion, production and destruction of charge species under the influence of the electric field. The electrostatics equations include the Poisson equation. 

The Poisson equation, which includes the space charge density as a source term, governs the electric potential \(V\):
\begin{equation}\label{e1}
\nabla^2V=-\frac{\rho}{\varepsilon}  \\
\end{equation}
where \(\rho\) is the net space charge density, and \(\epsilon\) is the air permittivity. 

The electric field intensity \(E\) is computed by taking the gradient of the electric potential, given as \( \vec{E} = - \nabla V \). The net space charge density \(\rho\) is provided as:"

\begin{equation}\label{e2}
\rho=\rho_p-\rho_n-\rho_e =  \textbf{e} \cdot (c_p - c_n -c_e ) 
\end{equation}
where \(c\) is the number density of charge species, the subscripts \(e\), \(n\) and \(p\) refer to the electrons, positive ions and negative ions, respectively, and \(\textbf{e}\) is the electronic charge.

Within the fluid description of the discharge, the ionic particle density is characterized through a system of charge continuity equations based on the Eulerian formulation \cite{kourtzanidis2020self}. The transport equations incorporate flux terms that consider both drift influenced by a Poissonian electric field and diffusion. The source terms encompass a range of chemical reactions, including ion-ion recombination, electron attachment and detachment, and ionization resulting from electron impacts. The following are the charge continuity equations:

\begin{equation}\label{e5}
\begin{cases}

\frac{\partial c_e}{\partial t}+\nabla\cdot\left(- c_e\mu_e\vec{E}-D_e\nabla c_e\right)= S_e \\

\frac{\partial c_p}{\partial t}+\nabla\cdot\left( c_p\mu_p\vec{E}-D_p\nabla c_p\right)= S_p \\ 

\frac{\partial c_n}{\partial t}+\nabla\cdot\left(- c_n\mu_n\vec{E}-D_n\nabla c_n\right)= S_n \\     
\end{cases}
\end{equation}
Here, \(t\) represents time, \( \vec{E}\) denotes the electric field, \(\mu\) represents the charge species' mobility, \(D\) their diffusivity, and \(S\) stands for the reaction terms. The mobilities of the positive and negative ions are equal to \(2.43\times{10}^{-4}\;{m^2}{\left(V\cdot s\right)}^{-1}\) and \(2.70\times{10}^{-4}\;{m^2}{\left(V\cdot s\right)}^{-1}\), respectively \cite{georghiou2005numerical}. The diffusion coefficient of the ions is equal to \(4.3\times{10}^{-6}\;{m^2}{s}^{-1}\) \cite{nishida2016numerical}. The calculation of the electron diffusion coefficient relies on the Einstein relationship \( \large{\sfrac{D_e}{\mu_e}} =  1 {V} \) \cite{nakai2021validity}. The definitions of the source reaction terms \(S\) are as follows:

\begin{equation}\label{eSo}
\begin{cases}

S_e = \alpha c_e | \mu_e \vec{E} |  -\beta_{pe} c_e c_p   -  \eta c_{e} | \mu_e \vec{E} |    \\

S_p = \alpha c_{e} | \mu_e \vec{E} |   - \beta_{pe} c_p c_e    - \beta_{np} c_p c_n  \\

S_n =    \eta  c_{e} | \mu_e \vec{E} |  - \beta_{np} c_p c_n  \\
\end{cases}
\end{equation}

The coefficients are defined as follows: \(\alpha \) for ionization,  \(\beta \) for recombination, and \( \eta \) for attachment, with subscripts \(e\), \(p\) and \(n\) representing electrons, and positive and negative ions, respectively. The recombination  coefficient \(\beta\) is equal to \(2.0\times{10}^{-13}\;{m^3}{s}^{-1}\). Electron mobility and diffusivity are expressed as \\ \(1.9163\times|\vec{E} |^{-0.25} \;{m^2}{\left(V\cdot s\right)}^{-1}\) and \(0.18 \; {m^2}{s}^{-1}\), respectively, and the ionization and attachment coefficients are expressed as \( 3.5\times{10}^{5} exp(-1.65 \times{10}^{7}/|\vec{E}|)  \;{m}^{-1}\) and \( 1.5\times{10}^{3} exp(-2.5 \times{10}^{6}/|\vec{E}|)  \;{m}^{-1}\), respectively \cite{dordizadeh2015numerical,shaygani2021dielectric}.

Seed electrons for self-sustained pulses are generated by imposing an electron flux boundary condition on the discharge electrodes as per the equation \( \Gamma_e =   \gamma  c_p \mu_p | \vec{E} |  \). Here \(\Gamma_e\) represents the electron flux and \(\gamma\) denotes the secondary electron emission coefficient \cite{nakai2020effect}. It shows how the seed electrons are produced due to the collision of positive ions onto the metal surface of the discharge electrode.

Among the charge species present within the domain, some participate in ionic processes such as neutralization or formation, while others exit the computational domain through zero convective flux boundary conditions. The simulation starts based on the assumption of charge neutrality, for the initial condition of charge densities. For the electrostatics equations, \(V_{ap}\) is assumed on the discharge electrode, and zero voltage on the passive electrode. The far-field air boundaries are assigned a zero normal component of the displacement vector. As for the species transport, the secondary electron emission flux is enforced at the discharge electrodes for electrons, zero charge for the negative ions, and zero convective flux for the positive ions. On the ground electrode, zero value is set for the positive ions, and zero flux for the negative ions and electrons. Remote boundaries have either zero convective flux or zero charge conditionally, for all species. A summary of the boundary conditions can be found in Table \ref{bcSync}.

\newcolumntype{L}{>{\raggedright\arraybackslash}X}
\begin{table}[htbp]
\relscale{0.7}
\begin{center}
\caption{Boundary conditions}

\begin{tabularx}{\linewidth}{L L L L}
\hline
\hline
\\ [-5pt] 
\textbf{Equation (variable)}& \textbf{Active electrode} & \textbf{Passive electrode} & \textbf{Remote boundary}
\\ [+2pt]
\hline
\\ [-5pt]

Poisson \((V, \Vec{E})\)  & \(V = V_{ap}\) &  \(V = 0\)    &\(\vec{n}\cdot(\varepsilon\vec{E})=0\) \\

Electrons \((c_e)\) &  \(\text{Electron flux}=\Gamma_e\)  & \(\vec{n}\cdot(\nabla c_e)=0\) & \[\begin{cases}
\vec{n}\cdot(\nabla c_e)=0  &  \vec{n}\cdot(- \mu_e \Vec{E}) > 0   \\
c_e=0   &  \vec{n}\cdot(- \mu_e \Vec{E}) \leq 0
\end{cases}\] \\

Positive ions \((c_p)\) & \(\vec{n}\cdot(\nabla c_p)=0\) & \( c_p=0\) &  \[\begin{cases}
\vec{n}\cdot(\nabla c_p)=0  &  \vec{n}\cdot( \mu_p \Vec{E}) > 0   \\
c_p=0   &  \vec{n}\cdot( \mu_p \Vec{E}) \leq 0
\end{cases}\]  \\

Negative ions \((c_n)\) & \( c_n=0\) & \(\vec{n}\cdot(\nabla c_n)=0\) &  \[\begin{cases}
\vec{n}\cdot(\nabla c_n)=0  &  \vec{n}\cdot(- \mu_e \Vec{E}) > 0   \\
c_n=0   &  \vec{n}\cdot(- \mu_n \Vec{E}) \leq 0
\end{cases}\] \\

\\ [-5pt]
\hline
\hline

\end{tabularx}
\label{bcSync}
\end{center}
\end{table}

\newpage

\subsection{Governing equations of the \(1D\)-axisymmetric model}

The \(1D\)-axisymmetric model consists of two separate discharge systems, each of which involves a pair of electrodes: one discharge and one ground. Since the individual systems are one-dimensional and coupling through physical dimensions is impossible, it is assumed that they are coupled mathematically through three 
quantities, space charges \(\rho\), electric fields \(\vec{E}\), and number densities \(c\).

The governing equations are similar to that of the \(2D\) model, with additional terms introduced as coupling terms. The terms are multiplied by three coupling parameters \(P_1\), \(P_2\), and \(P_3\), by which the strength of the couplings can be regulated. \(P_1\) introduces the influence of the space charge density of the other discharge system. Similarly, \(P_2\) introduces the influence of the number densities for the reaction terms, and \(P_3\) introduces the influence of the electric field. A parameter set to zero signifies the absence of coupling for the corresponding quantity of the other discharge system, while a parameter set to one indicates an equal influence of the quantity from the other discharge system, on par with self-influence. The parameters are listed in Table \ref{cop}.

The governing equations are given in (\ref{RhoCop}) to (\ref{S2Cop}). Each discharge system is governed by a set of electrostatics equations (\ref{RhoCop} - \ref{StCop}) and transport equations (\ref{TrCop} - \ref{S2Cop}). Subscripts \(1\) and \(2\) label the respective discharge systems.

\begin{equation}\label{RhoCop}
\begin{cases}
\rho_1=\rho_{1p}-\rho_{1n}-\rho_{1e} = \textbf{e} \cdot (c_{1p} - c_{1n} -c_{1e} ) 
\\
\rho_2=\rho_{2p}-\rho_{2n}-\rho_{2e} =  \textbf{e} \cdot (c_{2p} - c_{2n} -c_{2e} ) 
\end{cases}
\end{equation}

\begin{equation}\label{StCop}
\begin{cases}
\nabla \cdot  (\vec{E_1} - P_3 \vec{E_2}) =-\frac{\rho_1 + P_1 \cdot \rho_2 }{\epsilon} \\
\vec{E_{1}}  = - \nabla V_{1}  \\
\\
\nabla \cdot  (\vec{E_2} - P_3 \vec{E_1})  =-\frac{\rho_2 + P_1 \cdot \rho_1 }{\epsilon} \\
\vec{E_{2}} = - \nabla V_{2}  \\
\end{cases}
\end{equation}

\begin{equation}\label{TrCop}
\begin{cases}

\frac{\partial c_{1e}}{\partial t}+\nabla\cdot\left(- c_{1e}\mu_{1e}\vec{E_{1}}-D_{1e}\nabla c_{1e}\right)= S_{1e} \\

\frac{\partial c_{1p}}{\partial t}+\nabla\cdot\left( c_{1p}\mu_{1p}\vec{E_{1}}-D_{1p}\nabla c_{1p}\right)= S_{1p} \\  

\frac{\partial c_{1n}}{\partial t}+\nabla\cdot\left(- c_{1n}\mu_{1n}\vec{E_{1}}-D_{1n}\nabla c_{1n}\right)= S_{1n} \\     

\\

\frac{\partial c_{2e}}{\partial t}+\nabla\cdot\left(- c_{2e}\mu_{2e}\vec{E_{2}}-D_{2e}\nabla c_{2e}\right)= S_{2e} \\

\frac{\partial c_{2p}}{\partial t}+\nabla\cdot\left( c_{2p}\mu_{2p}\vec{E_{2}}-D_{2p}\nabla c_{2p}\right)= S_{2p} \\ 

\frac{\partial c_{2n}}{\partial t}+\nabla\cdot\left(- c_{2n}\mu_{2n}\vec{E_{2}}-D_{2n}\nabla c_{2n}\right)= S_{2n} \\     
 
\end{cases}
\end{equation}

In the context of our study, we consider two distinct systems: System \(\#1\) and System  \(\#2\). Each source \(S\) comprises two components, introduces as (\ref{SoCop}). The first pertains to the source from the primary discharge system (self), while the second component corresponds to the source from the other discharge system with the coupling parameter \(P_2\). When formulating the equations for each system, we introduce notation that captures the mutual influences between their respective quantities. Specifically, to address the impact of System \(\#2\)'s variables on System \(\#1\), we employ the notation \(S_{1-2}\). Conversely, for the effect of System \(\#1\)'s variables on System \(\#2\), we use the symbol \(S_{2-1}\). Furthermore, we denote the internal influences within each system using \(S_{1-1}\) and \(S_{2-2}\) to represent the self-influence of quantities within System \(\#1\) and System \(\#2\), respectively. This systematic notation allows us to succinctly derive the interplay and dependencies between the two systems, facilitating analysis of their dynamic interactions and behaviors. The terms are expanded in (\ref{S1Cop}) and (\ref{S2Cop}):

\begin{equation}\label{SoCop}
\begin{cases}

S_{1e} = S_{(1e-1e)} + P_2 \cdot S_{(1e-2e)}   

\\

S_{1p} = S_{(1p-1p)} + P_2 \cdot S_{(1p-2p)}   

\\

S_{1n} =   S_{(1n-1n)} + P_2 \cdot S_{(1n-2n)}  

\\
\\

S_{2e} = S_{(2e-2e)} + P_2 \cdot S_{(2e-1e)}  

\\

S_{2p} = S_{(2p-2p)} + P_2 \cdot S_{(2p-1p)}  

\\

S_{2n} =   S_{(2n-2n)} + P_2 \cdot S_{(2n-1n)}  

\\

\end{cases}
\end{equation}

\begin{equation}\label{S1Cop}
\begin{cases}

 S_{(1e-1e)} = \alpha_1 c_{1e} | \mu_{1e} \vec{E_1} |  -\beta_{pe} c_{1e} c_{1p}   -  \eta_1 c_{1e} | \mu_{1e} \vec{E_1} |    \\

 S_{(1p-1p)} = \alpha_1 c_{1e} | \mu_{1e} \vec{E_1} |   - \beta_{pe} c_{1p} c_{1e}    - \beta_{np} c_{1p} c_{1n}  \\

 S_{(1n-1n)} =    \eta_1  c_{1e} | \mu_{1e} \vec{E_1} |  - \beta_{np} c_{1p} c_{1n}  \\

\\

 S_{(2e-2e)} = \alpha_2 c_{2e} | \mu_{2e} \vec{E_2} |  -\beta_{pe} c_{2e} c_{2p}   -  \eta_2 c_{2e} | \mu_{2e} \vec{E_2} |    \\

 S_{(2p-2p)} = \alpha_2 c_{2e} | \mu_{2e} \vec{E_2} |   - \beta_{pe} c_{2p} c_{2e}    - \beta_{np} c_{2p} c_{2n}  \\

 S_{(2n-2n)} =    \eta_2  c_{2e} | \mu_{2e} \vec{E_2} |  - \beta_{np} c_{2p} c_{2n}  \\

\end{cases}
\end{equation}

\begin{equation}\label{S2Cop}
\begin{cases}

 S_{(1e-2e)} = \alpha_1 c_{2e} | \mu_{1e} \vec{E_1} |  -\beta_{pe} c_{1e} c_{2p} 
 -\beta_{pe} c_{2e} c_{1p} -  \eta_1 c_{2e} | \mu_{1e} \vec{E_1} |    \\

 S_{(1p-2p)} = \alpha_1 c_{2e} | \mu_{1e} \vec{E_1} |   - \beta_{pe} c_{1p} c_{2e} 
 - \beta_{pe} c_{2p} c_{1e} - \beta_{np} c_{1p} c_{2n} - \beta_{np} c_{2p} c_{1n} \\

 S_{(1n-2n)} =    \eta_1  c_{2e} | \mu_{1e} \vec{E_1} |  - \beta_{np} c_{1p} c_{2n}  - \beta_{np} c_{2p} c_{1n}\\

\\

 S_{(2e-1e)} = \alpha_2 c_{1e} | \mu_{2e} \vec{E_2} |  -\beta_{pe} c_{2e} c_{1p} -\beta_{pe} c_{1e} c_{2p}  -  \eta_2 c_{1e} | \mu_{2e} \vec{E_2} |    \\

 S_{(2p-1p)} = \alpha_2 c_{1e} | \mu_{2e} \vec{E_2} |   - \beta_{pe} c_{2p} c_{1e}  - \beta_{pe} c_{1p} c_{2e}    - \beta_{np} c_{2p} c_{1n} - \beta_{np} c_{1p} c_{2n} \\

 S_{(2n-1n)} =    \eta_2  c_{1e} | \mu_{2e} \vec{E_2} |  - \beta_{np} c_{2p} c_{1n} - \beta_{np} c_{1p} c_{2n}  \\

\end{cases}
\end{equation}

Similarly, this approach can be extended to establish the governing equations for three interconnected discharge systems. In this context, we focus solely on presenting the electrostatics equations to maintain conciseness, as incorporating the comprehensive transport equations along with their associated source terms would lead to a substantial increase in complexity and length. Therefore, for three discharge systems, we have:

\begin{equation}\label{Rho3sys}
\begin{cases}
\rho_1=\rho_{1p}-\rho_{1n}-\rho_{1e} =  \textbf{e} \cdot (c_{1p} - c_{1n} -c_{1e} ) 
\\
\rho_2=\rho_{2p}-\rho_{2n}-\rho_{2e} =  \textbf{e} \cdot (c_{2p} - c_{2n} -c_{2e} ) 
\\
\rho_3=\rho_{3p}-\rho_{3n}-\rho_{3e} =  \textbf{e} \cdot (c_{3p} - c_{3n} -c_{3e} ) 
\end{cases}
\end{equation}

\begin{equation}\label{St3sys}
\begin{cases}
\nabla \cdot  (\vec{E_1} - P_3 \vec{E_2} - P_3 \vec{E_3}) =-\frac{\rho_1 + P_1 \cdot \rho_2 + P_1 \cdot \rho_3}{\epsilon} \\
\vec{E_{1}}  = - \nabla V_{1}  \\
\\
\nabla \cdot  (\vec{E_2} - P_3 \vec{E_1} - P_3 \vec{E_3})  =-\frac{\rho_2 + P_1 \cdot \rho_1 + P_1 \cdot \rho_3}{\epsilon} \\
\vec{E_{2}} = - \nabla V_{2}  \\
\\
\nabla \cdot  (\vec{E_3} - P_3 \vec{E_1} - P_3 \vec{E_2})  =-\frac{\rho_3 + P_1 \cdot \rho_1  + P_1 \cdot \rho_2}{\epsilon} \\
\vec{E_{3}} = - \nabla V_{3}  \\
\end{cases}
\end{equation}

\begin{table}[htbp]
\caption{Coupling parameters}
\begin{center}
\begin{tabularx}{\linewidth}{L L} 

\hline
\hline     \\ [-5pt]                      

Coupling through the space charges in the electrostatics equations:  &
 \hspace{1cm}  \(   0 \leq P_1 \leq 1 \) \\
 
\hline     \\ [-5pt] 

Coupling through the number densities in the transport equations:  &
\hspace{1cm}  \(   0 \leq P_2 \leq 1 \)  \\

\hline     \\ [-5pt]

Coupling through the field effects in the electrostatics equations: &

\hspace{1cm}  \(   0 \leq P_3 \leq 1 \)  \\

    \\ [-5pt]

\hline     
\hline
\end{tabularx}
\label{cop}
\end{center}
\end{table}

\section{Simulation results}

\subsection{Numerical model}
The commercial FEM software COMSOL 6.1 was used to simulate the above model. The Poisson and Laplace equations with the second-order accurate interpolation scheme were solved in the Electrostatics module. The Laplacian field is calculated for obtaining the discharge currents. The charge continuity equations with the first-order scheme were solved in the Transport of Diluted Species module. In modeling discharge pulses, the continuity equations are highly advective and primarily characterized by the first derivatives of the species concentrations. Therefore, a first-order discretization scheme was used and the streamline diffusion method was adopted for stabilization. One should note that the Poisson equations of the \(1D\)-axisymmetric cases were modeled in the Coefficient Form PDE module, because of the additional terms containing parameters \(P_3\).

In the \(2D\) model, the computational domain underwent discretization using an unstructured triangular grid with around \(200 000\) elements, resulting in approximately \(1\) million degrees of freedom. Finer elements were employed near the discharge tips, especially within the ionization regions adjacent to each discharge tip. This fine discretization began at approximately \SI{500}{nm} and extended further away from both the discharge regions and the ground electrode. Iterative mesh refinement approach was adopted to ensure grid in-dependency of the simulation results. The same technique was also adopted for the \(1D\)-axisymmetric models while using a very fine mesh close to the discharge electrode, and gradually enlarging it near the grounded electrode. The implicit backward differentiation formulation (BDF) was employed, along with an adaptive time-stepping method, following spatial discretization. The equations were then solved in a fully-coupled manner within the time domain. For the \(2D\) and \(1D\)-axisymmetric problems, it takes around 50, and 2 CPU hours, respectively, which is calibrated for a system equipped with an Intel Core i7-6700 CPU running at 3.4
GHz.

\subsection{Model parameters}

Table \ref{tab1} provides a list of the parameters used in the discharge models. In the \(2D\) model, a parametric curve defines the profile of the blade electrodes: \(y=2 R t\), and \(x= R t^2 + l_1\), where \(R\) is called the thickness of the discharge electrode tip, \(y\) and \(x\) represents the spatial coordinates, and \(t\) serves as the parameter for the hyperbolic curve. In the \(1D\)-axisymmetric models, \(r_d\) and \(r_g\) are the radii of the discharge wires and grounded cylinders. The computational domains are a rectangular box for the \(2D\) model, and two and three \(1D\) lines in the \(1D\)-axisymmetric models. The voltage \(V_{ap}\) is applied to the discharge electrodes. A range of phase differences \(\phi\) are considered for the supplied voltages to the discharge electrodes.

\begin{table}[htbp]
\caption{The parameters for the models under investigation.}
\begin{center}
\begin{tabular}{l l}

\hline
\hline     \\ [-5pt]                            
\textbf{Parameter} & \textbf{Value}  \\ [+2pt] 

\hline    \\ [-5pt]

Air spacing between the electrodes, \(l_1\) & \SI{20}{mm}  \\
Distance between the electrodes, \(l_2\) & \(10\) to \SI{50}{mm}  \\
Width of the domain & \SI{80}{mm}  \\
Height of the domain & \SI{50}{mm}  \\ 
Blade electrode heights  & \SI{30}{mm} \\
Thickness of the hyperbolic profile & \SI{3}{mm} \\
Blade electrode DC voltages, \(V_{ap}\) &  \SI{-50}{kV}  \\ 
Relative permittivity of air, \(\epsilon_r\) & \(1.0006\)  \\ 
Wire electrode radii, \(r_d\) & \SI{200}{\mu m} \\
Cylinder electrode radii, \(r_g\) & \SI{10}{mm} \\  
Wire electrode DC voltages, \(V_{ap}\) &  \SI{-14}{kV}  \\
Phase delay, \(\phi\) & \(2\) to \SI{40}{\mu s}  \\ [+5pt]

\hline     
\hline
\end{tabular}
\label{tab1}
\end{center}
\end{table}

\newpage
\subsection{Pulse trains by the \(2D\) model}

\begin{figure}[htbp]
\centerline{\includegraphics[width=0.90\textwidth]{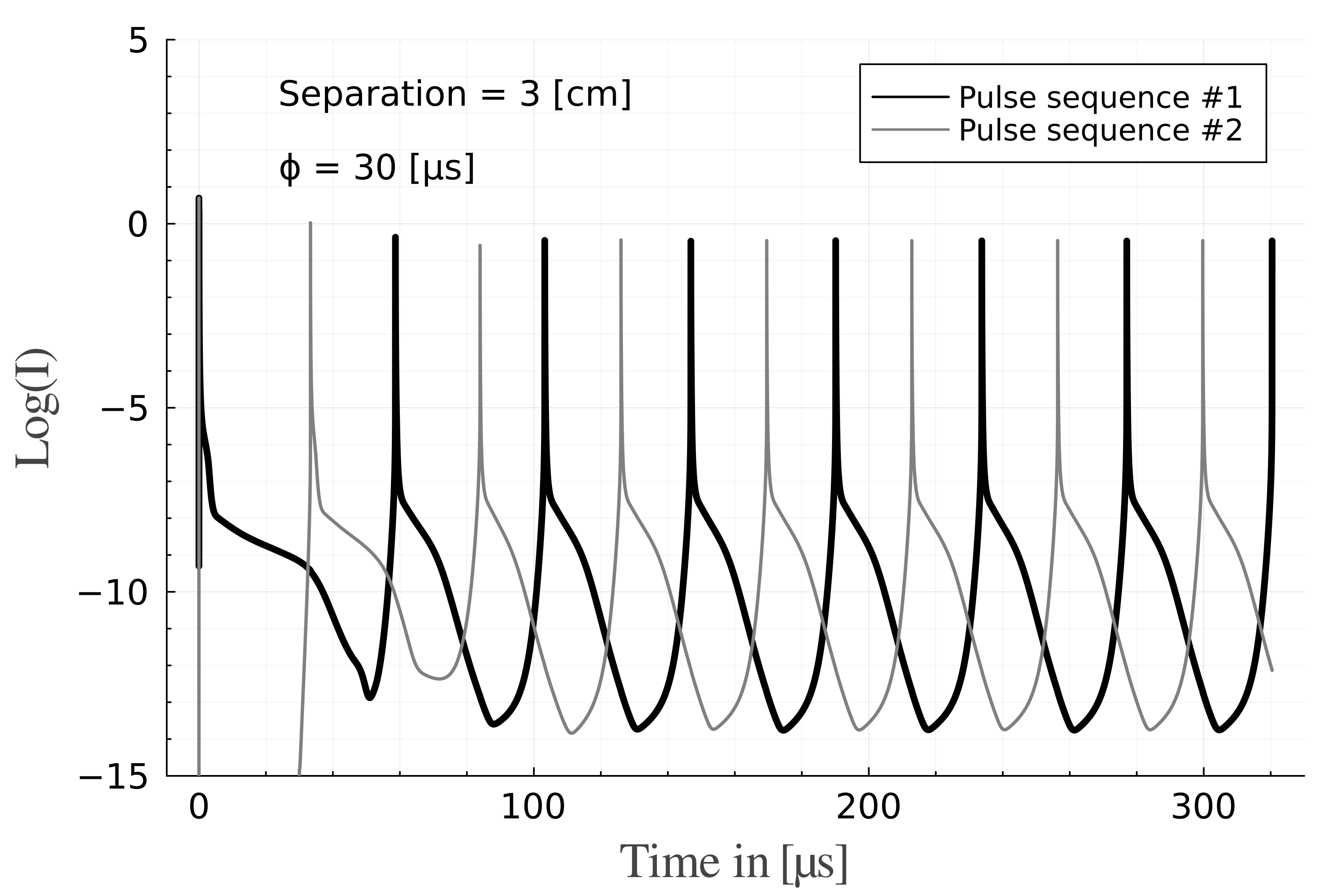}}
\caption{Anti-phase synchronization with \(l_2=\SI{3}{cm} \), \(\phi=\SI{30}{\mu s}\)}
\label{anti1d2}
\end{figure}

\begin{figure}[htbp]
\centerline{\includegraphics[width=0.90\textwidth]{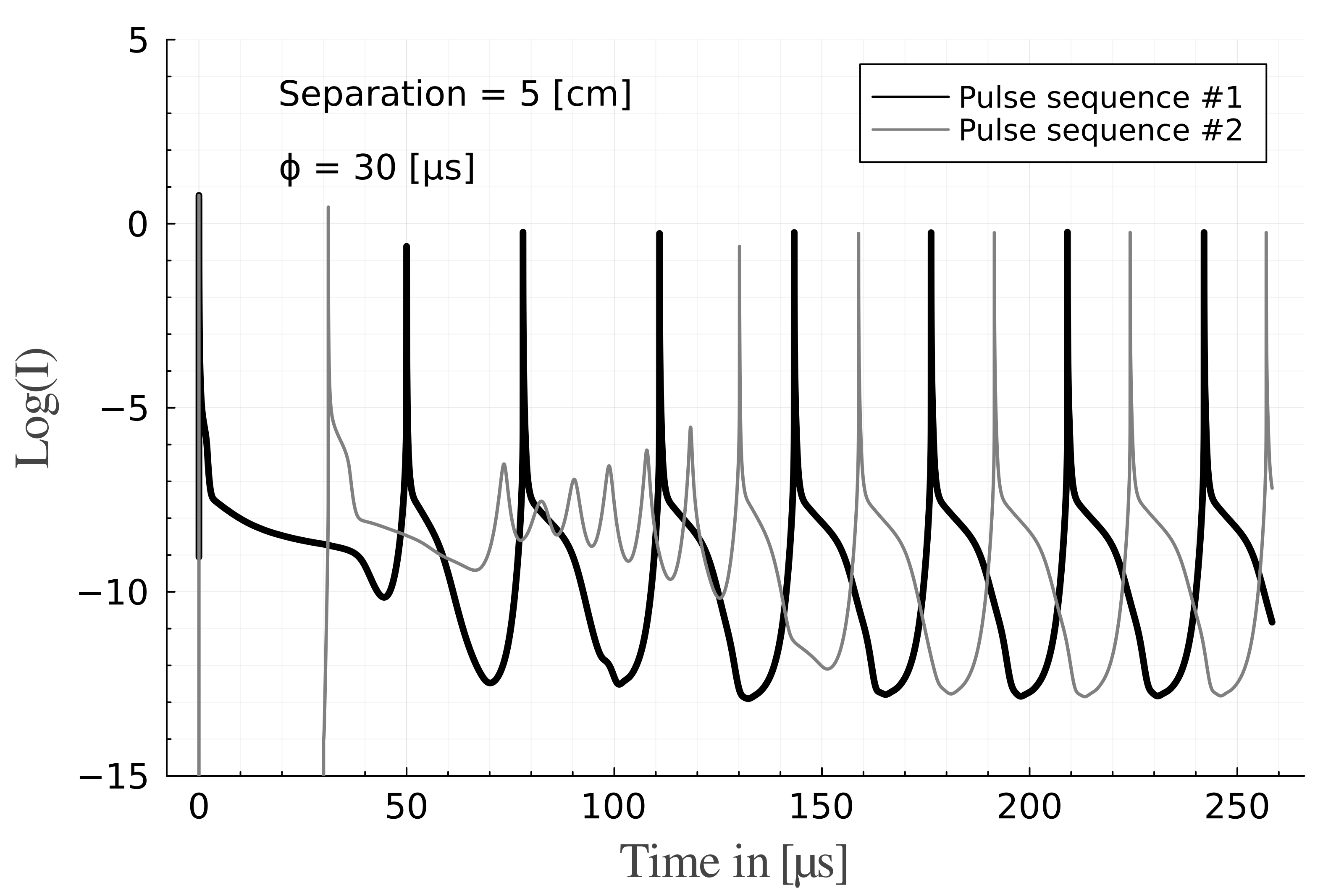}}
\caption{Anti-phase synchronization with \(l_2=\SI{5}{cm} \), \(\phi=\SI{30}{\mu s}\)}
\label{anti2d2}
\end{figure}

\begin{figure}[htbp]
\centerline{\includegraphics[width=0.90\textwidth]{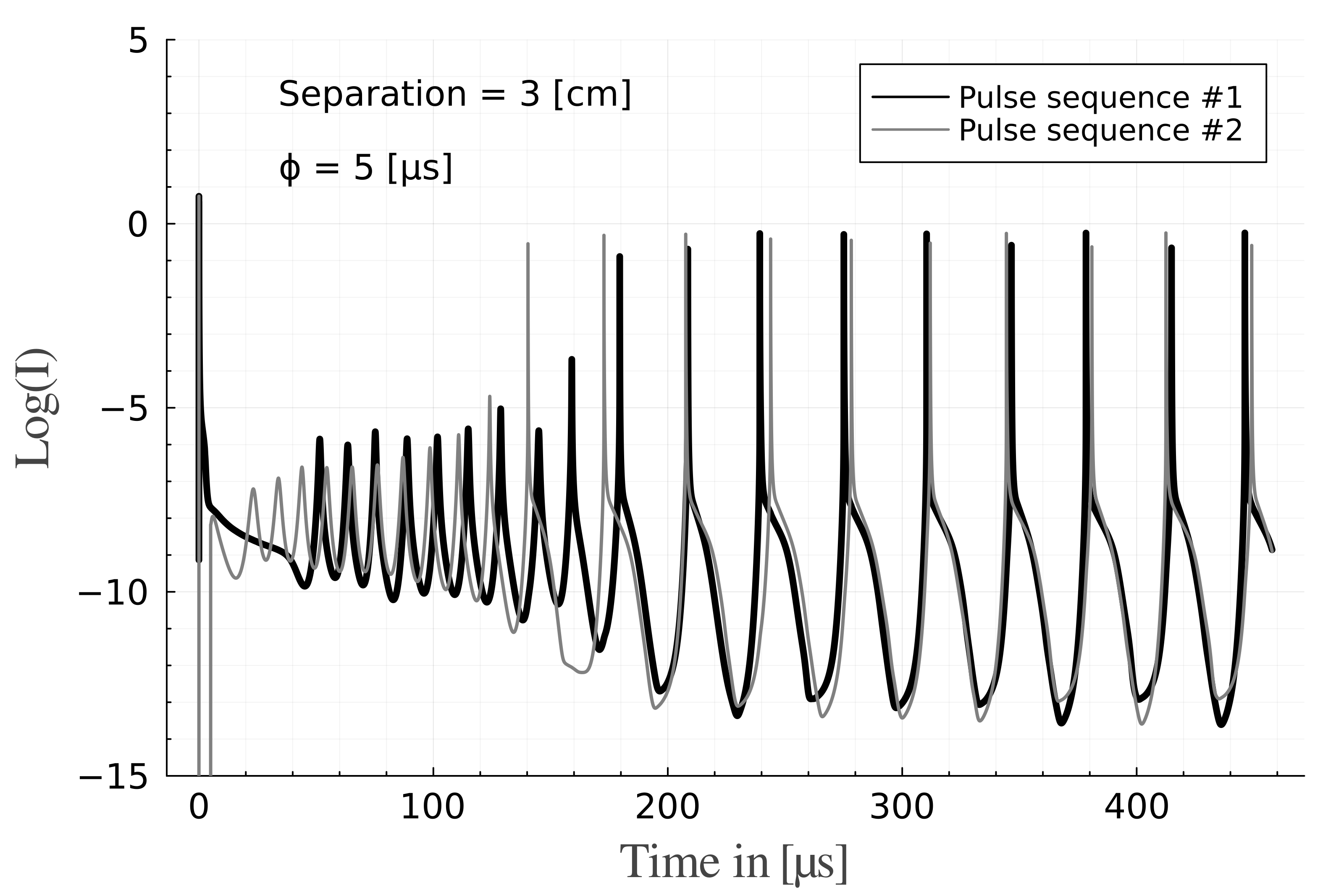}}
\caption{In-phase synchronization with \(l_2=\SI{3}{cm} \), \(\phi=\SI{5}{\mu s}\)}
\label{in1d2}
\end{figure}

\begin{figure}[htbp]
\centerline{\includegraphics[width=0.90\textwidth]{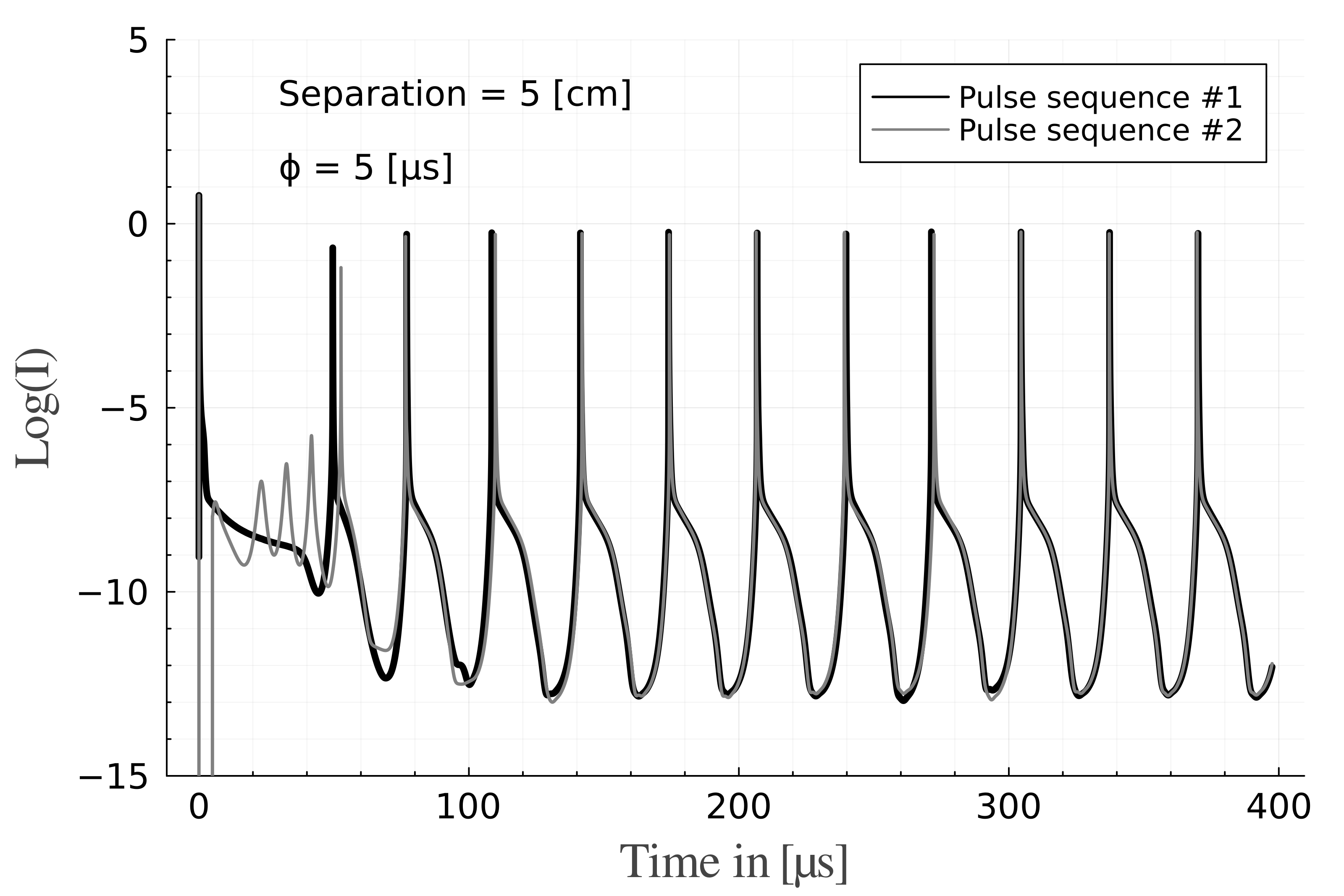}}
\caption{In-phase synchronization with \(l_2=\SI{5}{cm} \), \(\phi=\SI{5}{\mu s}\)}
\label{in2d2}
\end{figure}

\begin{figure}[htbp]
\centerline{\includegraphics[width=0.90\textwidth]{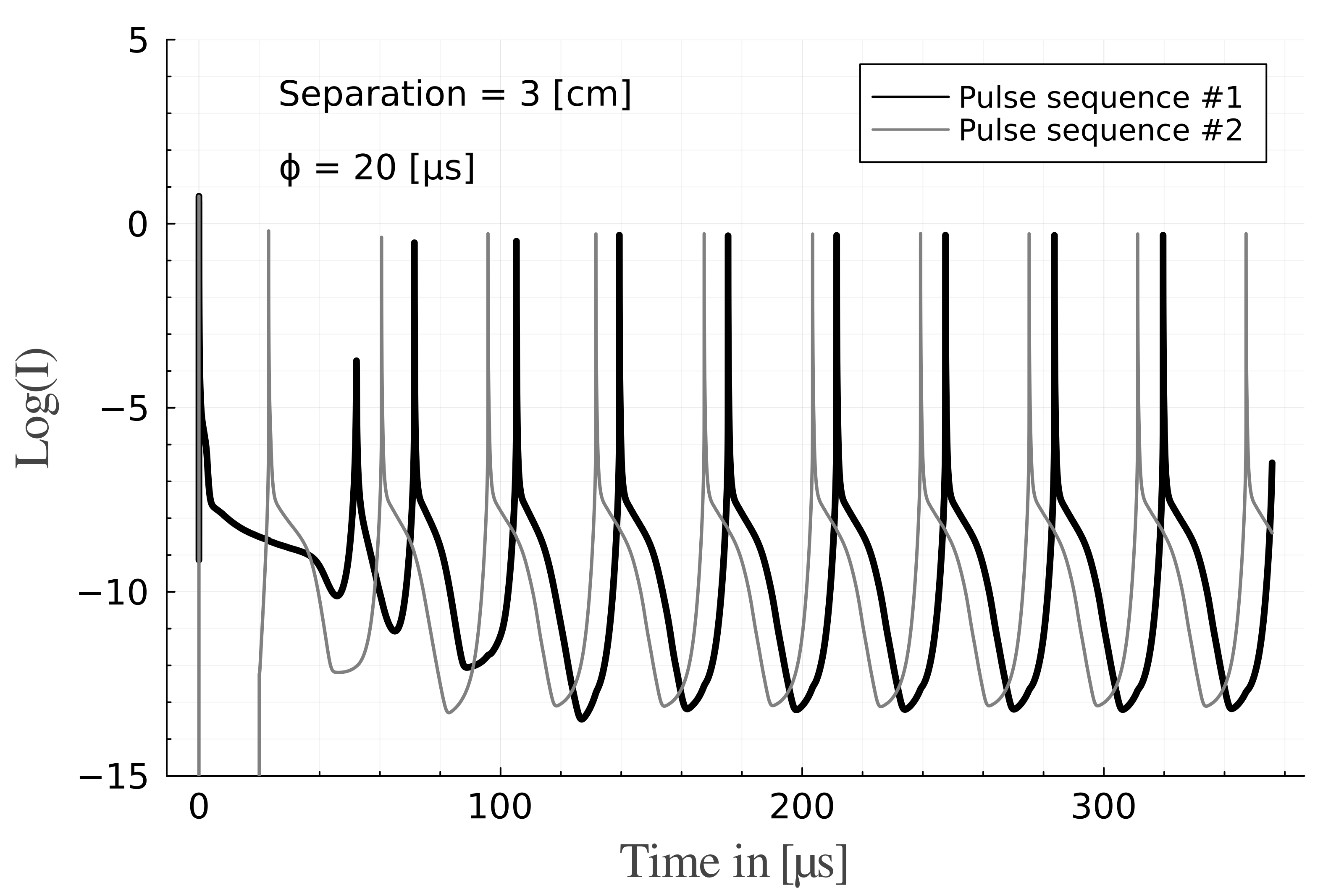}}
\caption{Out-of-phase synchronization with \(l_2=\SI{3}{cm} \), \(\phi=\SI{20}{\mu s}\)}
\label{out1d2}
\end{figure}

\begin{figure}[htbp]
\centerline{\includegraphics[width=0.90\textwidth]{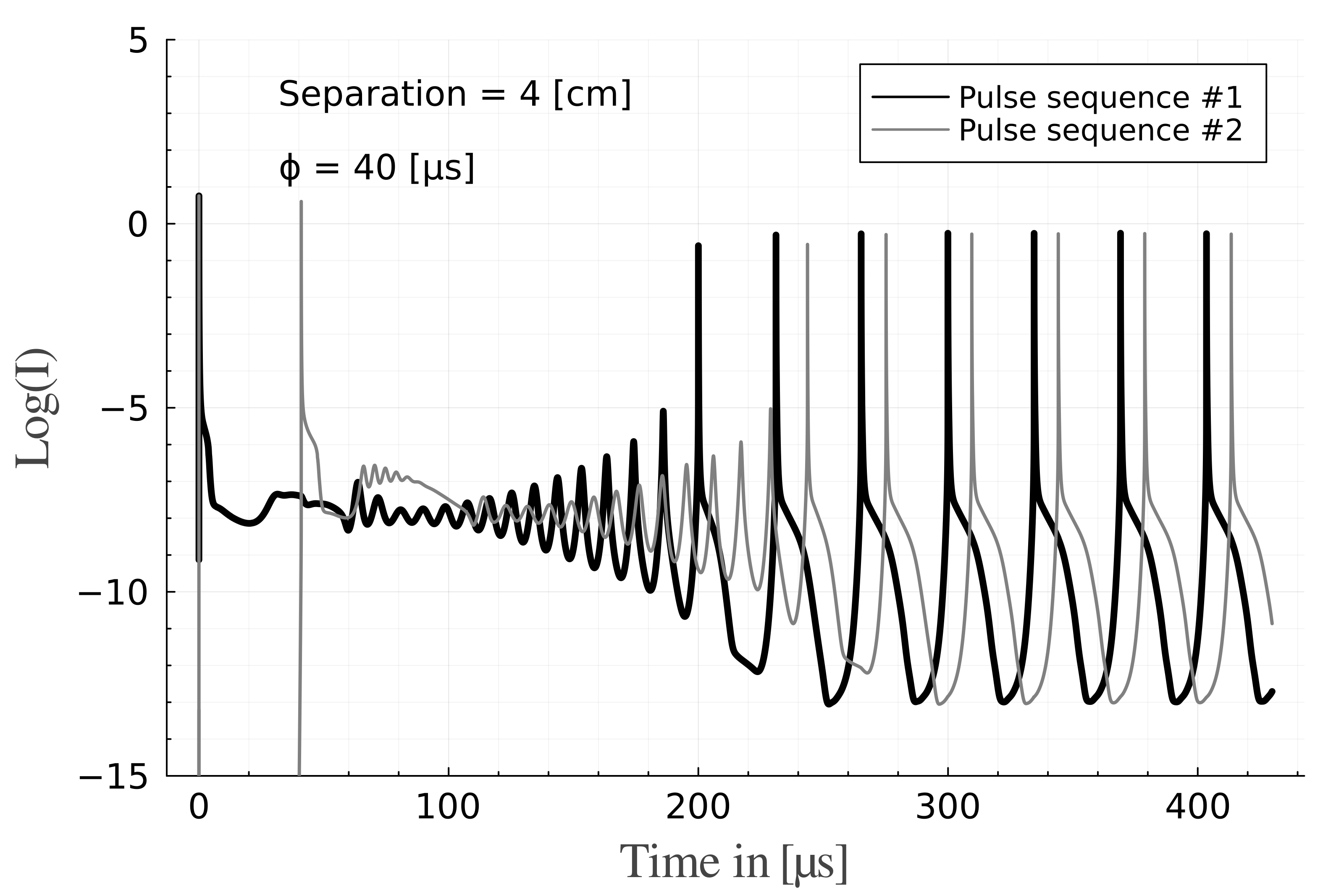}}
\caption{Out-of-phase synchronization with \(l_2=\SI{4}{cm} \), \(\phi=\SI{40}{\mu s}\)}
\label{out2d2}
\end{figure}

\begin{figure}[htbp]
\centerline{\includegraphics[width=0.90\textwidth]{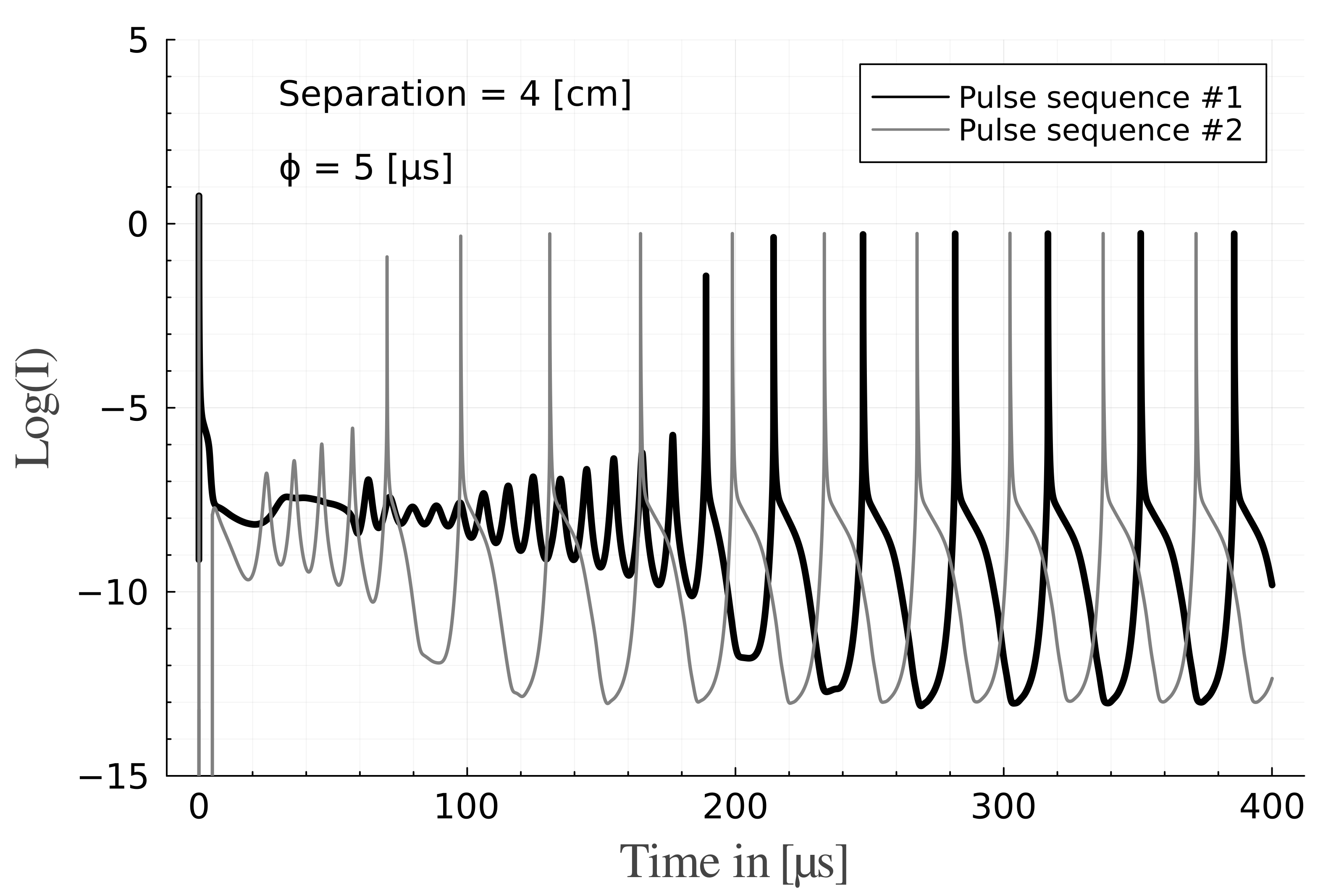}}
\caption{Out-of-phase synchronization with \(l_2=\SI{4}{cm} \), \(\phi=\SI{5}{\mu s}\)}
\label{out3d2}
\end{figure}

\begin{figure}[htbp]
\centerline{\includegraphics[width=0.90\textwidth]{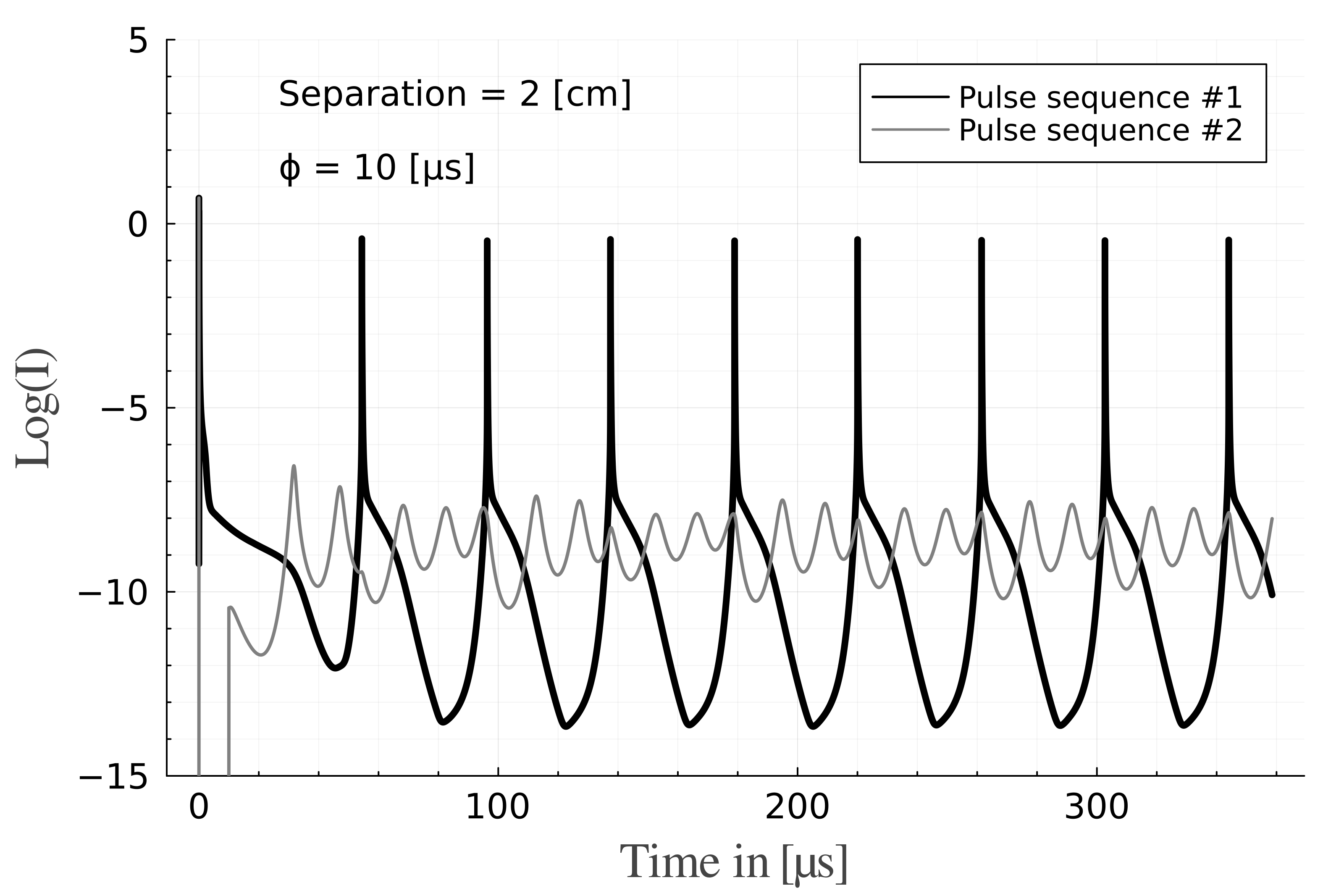}}
\caption{Quenching with \(l_2=\SI{2}{cm} \), \(\phi=\SI{10}{\mu s}\)}
\label{q1d2}
\end{figure}

The results of the \(2D\) simulations can be categorized into four distinct outcomes: anti-phase (Figs. \ref{anti1d2} and \ref{anti2d2}), in-phase (Figs. \ref{in1d2} and \ref{in2d2}), and out-of-phase synchronizations (Figs. \ref{out1d2}, \ref{out2d2}, and \ref{out3d2}), or quenching (Fig. \ref{q1d2}). All current pulse trains are displayed on a logarithmic scale, allowing the pulses to be more clearly distinguished and recognizable.

In the simplest and most basic models of synchronization, such as the
Kuramoto model \cite{kuramoto1975self,kuramoto1984chemical}, the final state of synchronization does not depend on the initial phase difference. This is not the case in some more complex systems such as the \(2D\) discharge systems studied here. If the interaction of the two discharge systems is not too weak (asynchronous) or too strong (quenching), the final synchronized state (phase) is dependent on the initial phase difference between the systems. If the initial phase difference between the two discharge systems is small, they fall into the in-phase synchronization. However, if the difference is closer to \(\frac{T}{2}\) - where \(T\) is the time period of the pulse trains when reached a quasi-steady state - they fall into the anti-phase synchronization. This suggests that the discharge systems are of nonlinear multi-stable type systems, since there exist multiple stable synchronized states.

Out-of-phase synchronization is also possible. All three kinds of synchronization occurring here are also dependent on the strength of the interaction, in other words, the distance between the two discharge electrodes. 

Quenching of discharges was also observed, particularly when the gap between the discharge electrodes is narrow. This phenomenon may occur for either both pulse trains or just one. When quenched, the pulses transform into high-frequency near-harmonic oscillations, reflecting a change in the regime of the discharges within the system.

\newpage
\subsection{Pulse trains by the \(1D\)-axisymmetric models}

\begin{figure}[htbp]
\centerline{\includegraphics[width=0.90\textwidth]{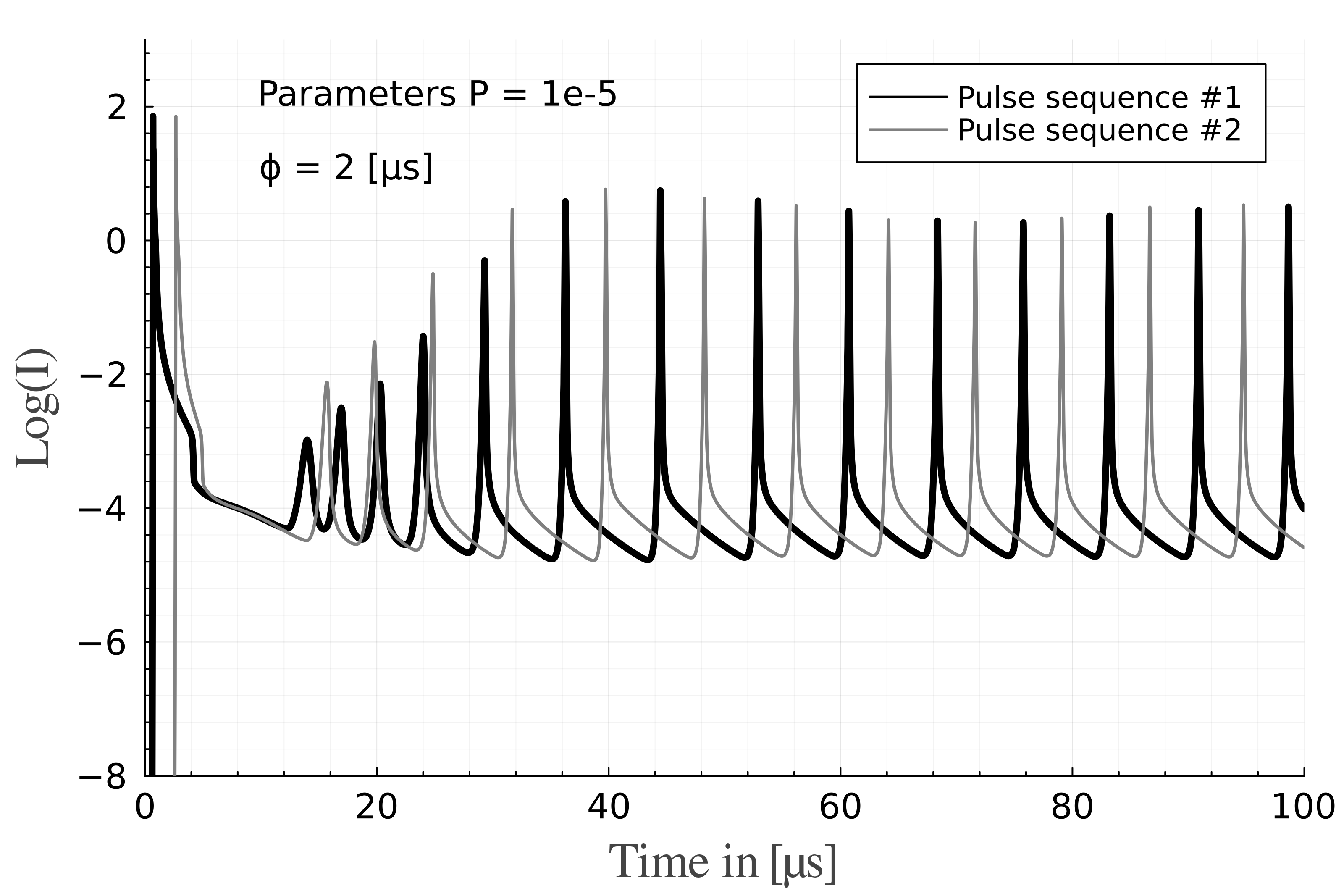}}
\caption{Anti-phase synchronization with \(P_i= 1e-5\), \(\phi=\SI{2}{\mu s}\)}
\label{anti2d1}
\end{figure}

\begin{figure}[htbp]
\centerline{\includegraphics[width=0.90\textwidth]{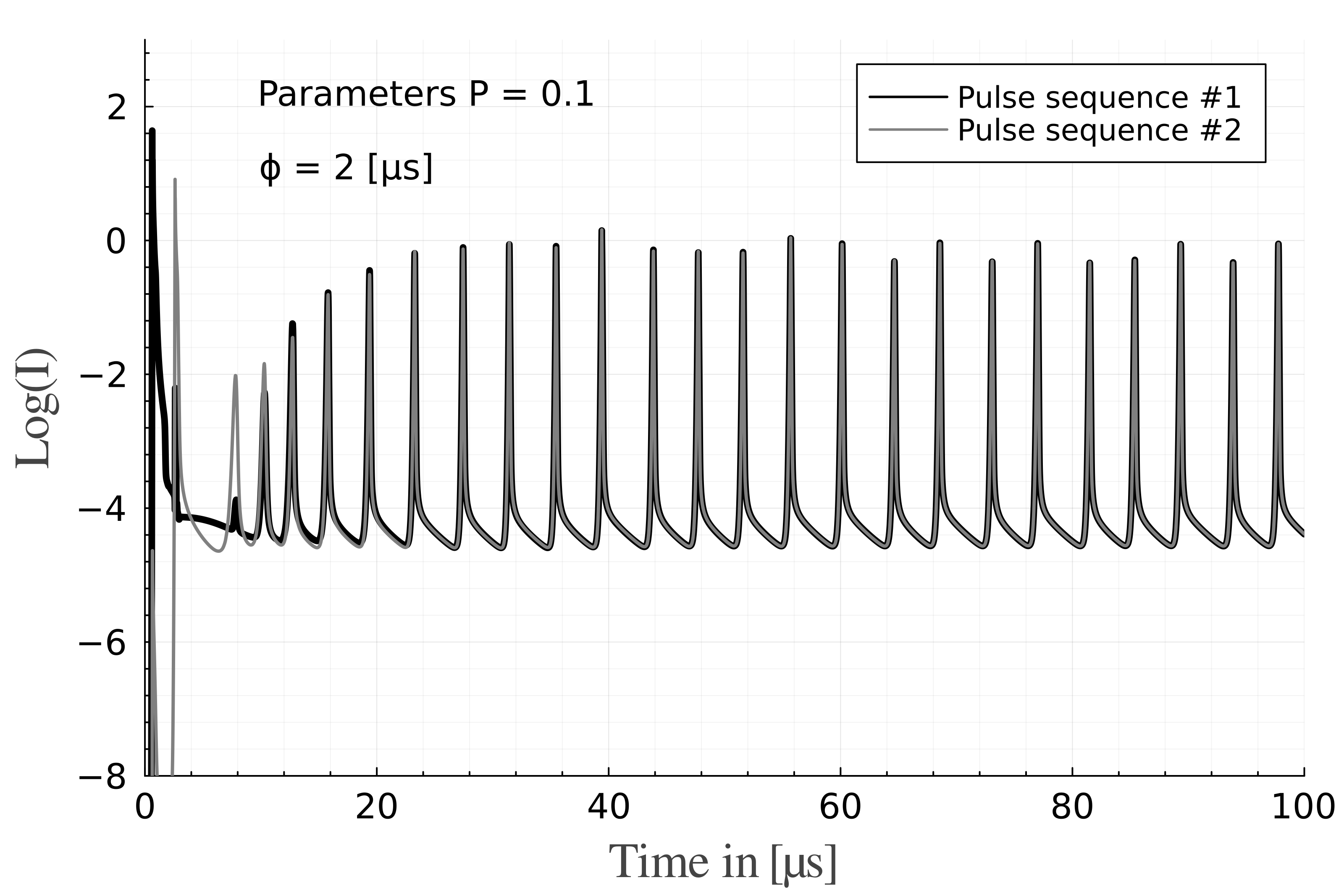}}
\caption{In-phase synchronization with \(P_i= 1e-2\), \(\phi=\SI{2}{\mu s}\)}
\label{in1d1}
\end{figure}

The results of the \(1D\)-axisymmetric simulations for the dual pair discharge system can be categorised into three outcomes: anti-phase (Fig. \ref{anti2d1}), in-phase (Fig. \ref{in1d1}), and out-of-phase synchronizations (Fig. \ref{out2d2}). The strength of the coupling is determined by \(P_i\) parameters, instead of the separation distance. For simplicity, all parameters were assumed to be equal. Using parameters as small as \(1e-6\) results in a synchronous behaviour.

The dynamics is less complex than that of the \(2D\) case, since the final state of synchronization does not depend on the initial phase difference. However, it is still dependent on the coupling strength. For small values of \(P_i\), anti- or out-of-phase synchronizations were observed. For large values, such as in Fig. \ref{in1d1} (\(P_i = 0.1\)) in-phase synchronization can be observed. For yet larger values, the solutions are divergent before quenching can be detected.

\begin{figure}[htbp]
\centerline{\includegraphics[width=0.90\textwidth]{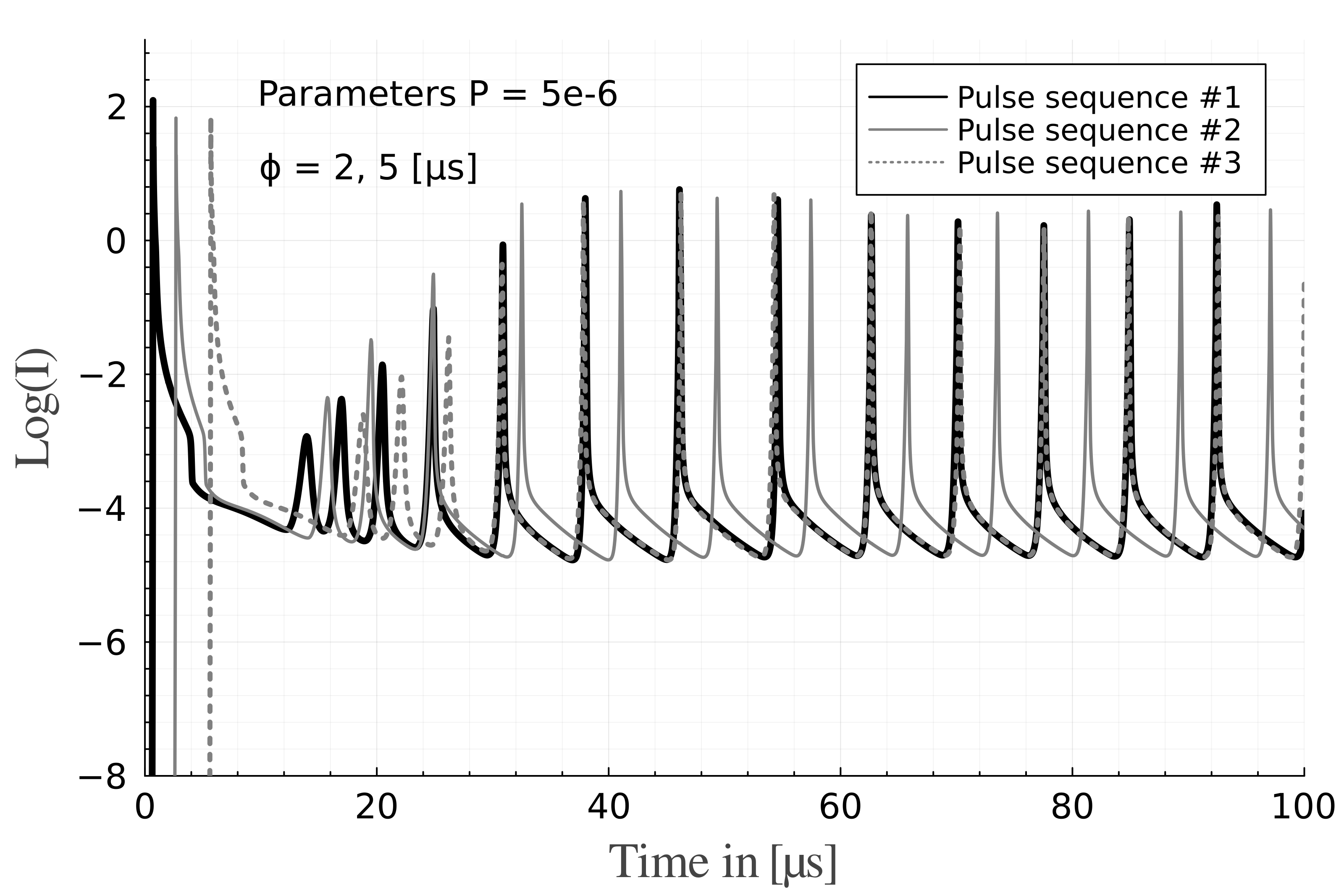}}
\caption{In and anti-phase synchronization with \(P_i= 1e-3\), \(\phi=2, \SI{5}{\mu s}\)}
\label{inanti2d1}
\end{figure}

The results of the \(1D\)-axisymmetric simulations for the triple pair discharge system are also shown.  Two pulse sequences are in-phase (\(\#1\) and \(\#2\)), and are anti-phase to \(\# 3\) (\ref{inanti2d1})

\begin{figure}[htbp]
\centerline{\includegraphics[width=0.90\textwidth]{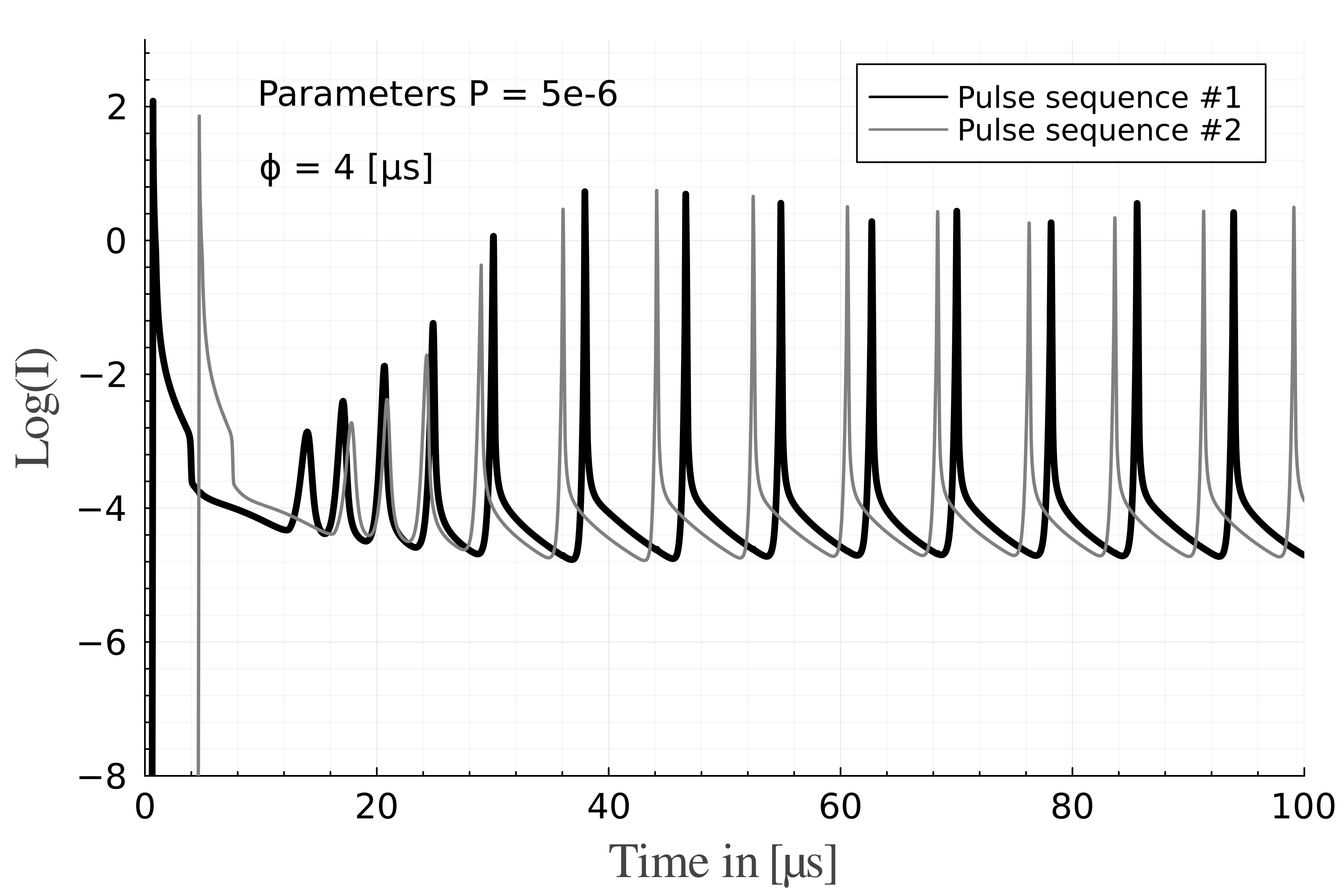}}
\caption{Out-of-phase synchronization with \(P_i= 5e-6\), \(\phi=\SI{4}{\mu s}\)}
\label{out2d1}
\end{figure}

\newpage
\section{Conclusions}
Self-synchronization has been observed in multi-electrode discharge systems operating in pulsed-mode negative corona discharges. Under appropriate conditions, pulsed-mode negative discharges emerge through application of high negative DC voltages to the discharge electrodes inducing self-sustained trains of current pulses, termed Trichel pulses. It has been found that with two or more discharge systems, each of which operating at Trichel pulse regime, weak interactions bring the systems into self-synchronization. 
These interactions result from the combined effects of electric fields, along with the transport and reactions of charged species.

The synchronization of Trichel pulses was first reported in an experimental study published in \cite{lama1973interaction}. Although the specific waveforms of the pulses were not captured, this investigation allowed the authors to identify synchronization by measuring currents. Despite this initial and significant finding, the subject has been largely overlooked since then, with no subsequent studies reporting similar observations. Our numerical study not only confirms the occurrence of synchronization, but also captures the pulse waveforms. Furthermore, it reveals that synchronization can occur in more than two discharge systems and provides a means to explore the origin of the synchronization, attributing it to weak interactions through the combined effects of space charge and electric field.

A three-species discharge model based on finite element method formulation was utilized, and distinct \(2D\) and \(1D\)-axisymmetric numerical models involving different electrode configurations were investigated. Numerical experiments demonstrate that under specific conditions, the pulse trains exhibit two synchronization modes: in-phase and anti-phase synchronization. The emergence of each mode hinges on factors, such as interaction strength, applied voltage, and various system parameters. Variations within these factors can result in additional outcomes, including out-of-phase synchronization, as well as scenarios involving near-harmonic oscillations and discharge quenching.

The \(1D\)-axisymmetric models effectively conceptualize weak interaction by incorporating coupling parameters for ionic reactions, electric field, and space charge. Notably, this approach not only offers a better computational efficiency compared to the \(2D\) model, but also permit the discrete examination of the influence of each of these factors – ionic reactions, electric field, and space charge – thereby providing enhanced insights into the nature of synchronization. Interestingly, our investigations have shown that independent coupling of these factors often leads to anti-phase synchronization. Our \(1D\)-axisymmetric models highlight that anti-phase synchronization is a predominant phenomenon. Specifically, for dual-pair discharge systems, the phase difference of the two pulse trains is \(180^\circ\), and for triple-pair configurations, the phase differences are \(0^\circ\) and  \(180^\circ \) (i.e. one pair anti-phase, the other in-phase). It is worth noting that, similar to the \(2D\) model, out-of-phase synchronization can also exist in \(1D\)-axisymmetric framework.

In our \(2D\) model, strong interactions – indicating closely spaced discharge electrodes – can lead to quenching of discharges. This quenching may occur for either both pulse trains or just one, resulting in their transformation into higher-frequency near-harmonic oscillations surpassing the original pulse frequencies. Conversely, when interactions are excessively weak, synchronization doesn't occure. Under appropriately mild interactions, current pulses can synchronize in either in-phase or anti-phase patterns. The occurrence of out-of-phase synchronization depends on factors such as supply voltage, distance between discharge electrodes, proximity to the grounded electrode, phase difference, and other system parameters.

Furthermore, our investigation points to the extension of synchronous behavior across a wider spectrum of electrode systems. Notably, it unveils the possibility of encountering random or chaotic pulsation modes under specific circumstances. In essence, this research significantly enriches our comprehension of self-synchronization within multi-electrode discharge systems, unveiling the intricate dynamics that underlie electric discharge phenomena. The implications of these findings extend across a broader spectrum of multi-system dynamics and may offer practical applications in diverse domains, including discharge-based space propulsion systems.

\hfill
\newpage
\bibliography{myrefACRF}

\end{document}